\documentclass[aps,10pt,twocolumn,showpacs,preprintnumbers,amsmath,amssymb,prl]{revtex4-1}

\usepackage{graphicx}

\newcommand{\ket}[1]{\left| #1 \right\rangle}
\renewcommand{\Re}{\operatorname{Re}}
\renewcommand{\Im}{\operatorname{Im}}
\newcommand{\expct}[2]{\left\langle #1 \right\rangle_{#2}}
\newcommand{\expcts}[2]{\langle #1 \rangle_{#2}}
\newcommand{\ph}{{\rm ph}}
\newcommand{\sgn}{{\rm sgn}}

\begin{document}

\title{Majorana qubit rotations in microwave cavities}

\author{Thomas~L.~Schmidt}
\author{Andreas~Nunnenkamp}
\author{Christoph~Bruder}

\affiliation{Department of Physics, University of Basel, Klingelbergstrasse 82, CH-4056 Basel, Switzerland}

\date{\today}

\begin{abstract}
    Majorana bound states have been proposed as building blocks for qubits on which certain operations can be performed in a topologically protected way using braiding. However, the set of these protected operations is not sufficient to realize universal quantum computing. We show that the electric field in a microwave cavity can induce Rabi oscillations between adjacent Majorana bound states. These oscillations can be used to implement an additional single-qubit gate. Supplemented with one braiding operation, this gate allows to perform arbitrary single-qubit operations.
\end{abstract}


\maketitle

The search for Majorana bound states (MBSs) \cite{kitaev01} in solid-state systems is currently receiving a lot of attention from theorists and experimentalists alike \cite{beenakker11,leijnse12,alicea12}. Shortly after the prediction that MBSs can be realized in semiconductor nanowires with strong Rashba spin-orbit coupling \cite{sato09,lutchyn10,oreg10}, various experiments have indeed reported signatures of Majorana fermions in such systems, which are currently being scrutinized \cite{mourik12,rokhinson12,deng12,das12}. MBSs appear at the phase boundaries between topologically trivial and nontrivial sections of the nanowire. The latter can be created in the presence of a proximity-induced superconducting gap $\Delta$ and a magnetic field $B$: if $\mu_0$ denotes the (position-dependent) chemical potential, the wire is in the topologically nontrivial (trivial) phase in regions where $\mu_0^2 < B^2 - \Delta^2$ ($\mu_0^2 > B^2 - \Delta^2$) \cite{oreg10}.

MBSs are interesting in their own right because they constitute the simplest quasiparticles with non-Abelian exchange statistics. They have also been suggested as resources for topological quantum computing \cite{nayak08} where the fact that a single qubit can be encoded in two or four spatially separated Majorana fermions offers protection from some common sources of decoherence.

Current proposals for topological quantum computing are striving to overcome two obstacles: First, it has been shown that the protection of MBSs from decoherence in realistic systems is less than ideal \cite{goldstein11,budich12a,rainis12}. Indeed, quasiparticle poisoning is probably a ubiquitous source of decoherence in all proposed setups to date. If one assumes that this problem can be overcome, a second, more fundamental difficulty arises: topologically protected operations on Majorana fermions, i.e., braiding, are not sufficient to achieve universal quantum computing \cite{nayak08}. They have to be complemented by other single-qubit and two-qubit gates -- as well as read-out operations -- which will still have to be performed in a topologically unprotected way. There have been proposals on how to perform such operations, e.g., in $p$-wave superfluids \cite{ohmi10,zhang07} or using conventional superconducting qubits \cite{hassler11}.

In this paper, we will focus on the second problem. To find a way to perform operations on a Majorana qubit in a topologically unprotected, but minimally invasive way, we will investigate two MBSs which are coupled via a gapped system of length $L$. In setups involving semiconductor nanowires, such a topologically trivial gapped region can be created using appropriate gating \cite{oreg10,mourik12}. Virtual cotunneling processes lead to an energy splitting between the two MBSs which is proportional to $e^{-L/\xi}$, where $\xi(\mu)$ is a model-dependent decay length which depends on the band gap $|\mu|$ and increases for $\mu \to 0$. We shall show that by coupling the MBSs to an electric field, it becomes possible to tune $\xi$, thus allowing for a coherent manipulation of the MBSs.

To see the benefits of such an endeavor, let us consider a popular proposal for a Majorana-based qubit. Four Majorana modes $\gamma_{1,2,3,4}$, which satisfy $\gamma_j^\dag = \gamma_j$ and $\{\gamma_j, \gamma_k\} = 2 \delta_{jk}$, can be used to define two Dirac fermion operators $\psi_L = (\gamma_1 + i\gamma_2)/2$ and $\psi_R = (\gamma_3 + i \gamma_4)/2$. Since the topological protection relies on a conserved fermion parity, the computational basis should contain states with the same parity. Therefore, one can use
\begin{align}\label{eq:eom_qubit}
    \ket{\downarrow} = \psi_L^\dag \ket{0},\quad
    \ket{\uparrow} =  \psi_R^\dag \ket{0}
\end{align}
as the two logical states of the qubit. Certain topologically protected single-qubit gates can be realized by braiding. For instance, exchanging the positions of the MBSs $\gamma_1$ and $\gamma_2$ ($\gamma_3$ and $\gamma_4$), which can be realized in semiconductor-nanowire based setups using $T$-shaped junctions \cite{alicea11}, corresponds to the transformations
\begin{align}
    U_{12} = \exp\left( \frac{i \pi}{4} \sigma_z \right),\quad
    U_{34} = \exp\left( - \frac{i \pi}{4} \sigma_z \right),
\end{align}
respectively, where $\sigma_z$ is a Pauli matrix in the basis $\{ \ket{\uparrow}, \ket{\downarrow} \}$. Similarly, exchanging $\gamma_2$ and $\gamma_3$ corresponds to $U_{23} = \exp\left(i \pi \sigma_x/4 \right)$. However, it is easy to see that these three operations are not sufficient to reach arbitrary points on the Bloch sphere spanned by all normalized linear combinations of $\ket{\uparrow}$ and $\ket{\downarrow}$.

In order to perform arbitrary single-qubit rotations, one needs to supplement this set of gates by topologically unprotected operations. If the MBSs $\gamma_2$ and $\gamma_3$ are subject to a coupling Hamiltonian
\begin{align}\label{eq:V}
    V = \frac{i \varepsilon}{2} \gamma_2 \gamma_3
\end{align}
for a certain time $t$, it corresponds to the following unitary transformation of the qubit state
\begin{align}\label{eq:Uepsilon}
    U_\varepsilon(t) = \exp\left( - \frac{i \varepsilon t}{2} \sigma_x \right).
\end{align}
Arbitrary points on the Bloch sphere can now be reached by combining this operation with braiding. For instance, the action of $U_\varepsilon(t_1) U_{12} U_\varepsilon(t_2)$ on the initial state $\ket{\uparrow}$ produces a final state with components,
\begin{align}
    \expct{\sigma_x}{} &= -\sin(\varepsilon t_2), \notag \\
    \expct{\sigma_y}{} &= -\sin(\varepsilon t_1) \cos(\varepsilon t_2),  \notag \\
    \expct{\sigma_z}{} &= \cos(\varepsilon t_1)\cos(\varepsilon t_2).
\end{align}
The vector $( \expct{\sigma_x}{}, \expct{\sigma_y}{}, \expct{\sigma_z}{})$ indeed covers the entire Bloch sphere for $\varepsilon t_2 \in [-\pi/2, \pi/2]$ and $\varepsilon t_1 \in [0, 2\pi]$. The same is true for all initial states on the Bloch sphere, so one can realize universal single-qubit rotations using the operator $U_\varepsilon(t)$ complemented by the braiding operation $U_{12}$. Therefore, our aim is to find a physical mechanism to realize the coupling Hamiltonian (\ref{eq:V}) in a controllable way.

\begin{figure}[t]
    \includegraphics[width=\columnwidth]{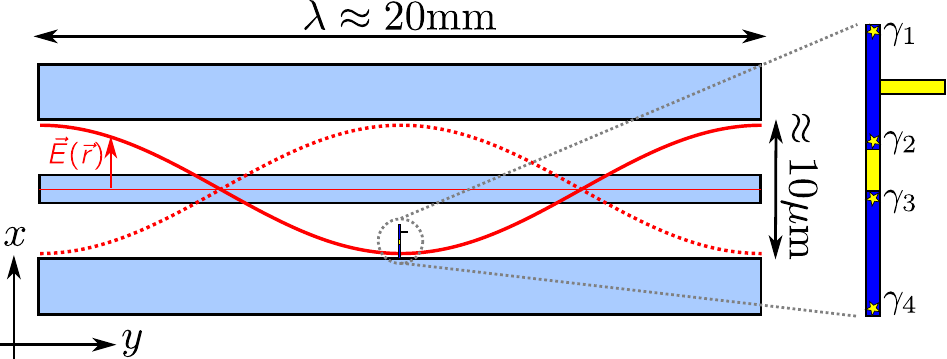}
    \medskip\\
    \includegraphics[width=\columnwidth]{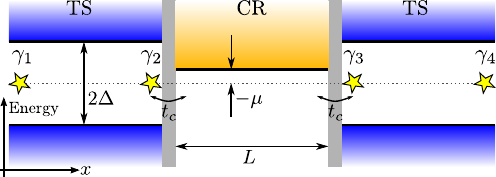}
    \centering
    \caption{(Color online) \emph{Upper panel:} A semiconductor nanowire (along the $x$ axis) hosting Majorana fermions is embedded in a microwave stripline cavity (along the $y$ axis). The red lines show the amplitude of the electric field $\vec{E}(\vec{r})$. Dark blue (light yellow) sections of the wire indicate topologically nontrivial (trivial) regions. MBSs (stars) exist at the edges of nontrivial (topological superconductor, TS) regions. The MBSs $\gamma_1$ and $\gamma_2$ can be braided using a $T$-junction \cite{alicea11}. \emph{Lower panel:} Band structure of the individual sections of the wire. The four MBSs $\gamma_{1,2,3,4}$ encode one logical qubit. The central MBSs $\gamma_{2}$ and $\gamma_3$ are tunnel-coupled ($t_c$) to a topologically trivial, gapped central region (CR, light yellow) with length $L$. All energies are small compared to the induced gap $\Delta$.}
    \label{fig:Schematic}
\end{figure}

Our starting point is the minimal system which contains the two MBSs $\gamma_2$ and $\gamma_3$, which are coupled indirectly by tunneling to a common, gapped central region (CR). The system is depicted schematically in Fig.~\ref{fig:Schematic}. It can be described by the Hamiltonian
\begin{align}\label{eq:H0}
    H_0 &= \sum_k \epsilon(k) d^\dag_k d_k \notag \\
    &- \frac{t_c}{2} \gamma_{2} [d(0) - d^\dag(0)] - \frac{i t_c}{2} [d(L) + d^\dag(L)] \gamma_{3}.
\end{align}
We assume that the generic single-particle spectrum of the CR is $\epsilon(k) = k^2/(2m) - \mu$, where $\mu < 0$. Here, $k$ is the wave number which labels the eigenmodes $d_k$, and $m > 0$ is an effective mass determined by the band curvature. The CR has a length $L$, and tunneling (with amplitude $t_c$) from and to the MBSs occurs at $x=0$ and $x=L$. In order to impose the proper boundary conditions, we use the conventional ``unfolding'' transformation \cite{fabrizio95} to map the CR onto a chiral system with length $2L$ containing only right-movers. One can then impose periodic boundary conditions on this doubled system, which leads to the momentum quantization $k = \pi n/L$, where $n \geq 0$. The fermion operators in real space are given by $d(x) = (1/\sqrt{2L}) \sum_k e^{i k x} d_k$.

The Hamiltonian $H_0$ is quadratic and can easily be solved exactly. One finds that tunneling leads to a nonzero overlap between $\gamma_2$ and $\gamma_3$. For small $t_c$, the resulting level splitting due to virtual cotunneling processes is
\begin{align}
    \varepsilon = \frac{t_c^2}{2 |\mu| \xi} e^{-L/\xi},
\end{align}
with the decay length $\xi = (2 m |\mu|)^{-1/2}$. At energies small compared to $|\mu|$, the effective coupling between the MBSs can be described by the Hamiltonian (\ref{eq:V}). Therefore, the operation $U_\varepsilon(t)$ could in principle be realized by tuning the chemical potential $\mu$ time-dependently. However, it is unlikely that this can be done on short timescales and accurately enough to realize qubit rotations reliably. Moreover, tuning $|\mu|$ to small values is undesirable. Once real tunneling processes become possible ($\mu > 0$), the Majorana qubit will quickly decohere \cite{budich12a}.

A possible way to realize a coupling between MBSs in a more controllable fashion is to use microwave photons to activate the tunnel process. Indeed, the induced gaps in recent experiments are in the microwave regime \cite{mourik12}. Moreover, hybrid structures involving semiconductor nanostructures and microwave cavities have recently been realized experimentally \cite{frey12}. It may at first be surprising that Majorana fermions, which are by definition uncharged, should be susceptible to electromagnetic radiation. However, today's solid-state versions are superpositions of particles and holes, and are chargeless only on average. The existence of a coupling of MBSs to a vector potential $A(x)$ can be derived explicitly by using the minimal-coupling substitution $p \to p - e A(x)$ in the Hamiltonian for semiconductor-nanowire based proposals \cite{lutchyn10,oreg10}. A schematic picture of a semiconductor nanowire embedded in a microwave cavity is shown in the upper panel of Fig.~\ref{fig:Schematic}.

Therefore, we place the previous system in a microwave cavity and describe it using the Hamiltonian,
\begin{align}\label{eq:H}
    H &= \sum_k \epsilon(k) d^\dag_k d_k - g_c (a + a^\dag) \gamma_{2} [d(0) - d^\dag(0)] \notag \\
    &- i g_c (a + a^\dag) [d(L) + d^\dag(L)] \gamma_{3} + H_\ph.
\end{align}
In contrast to Eq.~(\ref{eq:H0}), a tunnel process between the CR and the MBSs is now accompanied by the emission or absorption of a photon. The cavity degrees of freedom are described by the Hamiltonian $H_\ph = \Omega a^\dag a + H_{\rm bath}$, where $\Omega$ is the resonance frequency and $H_{\rm bath}$ accounts for driving and damping of the cavity. A similar form of activated tunneling into MBSs was investigated previously in Ref.~\cite{walter11b}.

\begin{figure}[t]
    \includegraphics[width=\columnwidth]{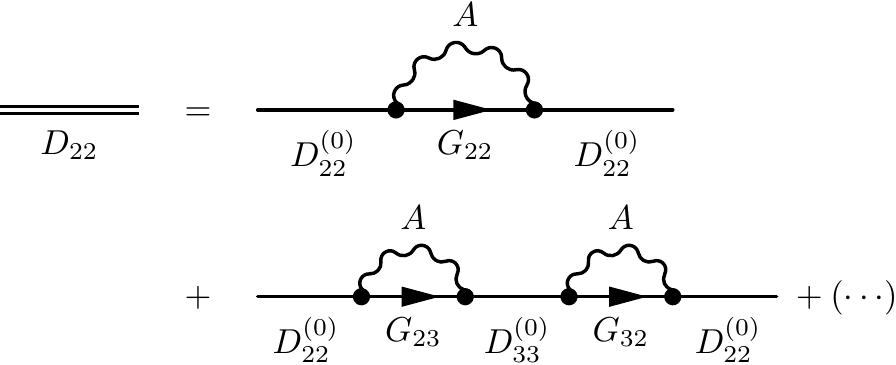}
    \centering
    \caption{Feynman diagram for the lowest-order terms of the Dyson equation (\ref{eq:pert_D2}). $D_{ij}$, $G_{ij}$, and $A$ denote the Green's functions of the Majorana bound states, the central region, and the photon, respectively.}
    \label{fig:Feynman}
\end{figure}

In order to determine the effective coupling between the MBSs due to the electron-photon interactions, we calculate the time-ordered Green's functions,
\begin{align}
    D_{ij}(t) = -i \expct{T \gamma_i(t) \gamma_j(0)}{},
\end{align}
where $T$ denotes the time-ordering operator and $i,j \in \{2,3\}$. Since the Hamiltonian $H$ is not exactly solvable, we use a perturbative approach valid at small coupling strength $g_c$ and perform a resummation of Fock-type diagrams, see Fig.~\ref{fig:Feynman}. The result for $D_{22}(\omega)$ is
\begin{align}\label{eq:pert_D2}
    D_{22} &= D^{(0)}_{22} +
    D^{(0)}_{22} \Sigma_{22} D_{22}
+
    D^{(0)}_{22} \Sigma_{23} D^{(1)}_{33} \Sigma_{32} D_{22},
\end{align}
where $[D^{(1)}_{33}]^{-1} = [D^{(0)}_{33}]^{-1} - \Sigma_{33}$. The unperturbed Majorana Green's functions are trivial because for $g_c = 0$, $\gamma_{2,3}$ are time-independent, i.e., $D^{(0)}_{22}(t) = D^{(0)}_{33}(t) = -i \sgn(t)$. The self-energy is due to tunneling with absorption or emission of a photon, $\Sigma_{ij}(t) = i g_c^2 A(t) G_{ij}(t)$, where $A(t)$ and $G_{ij}(t)$ are Green's functions of the photon and the CR, respectively, which will be discussed in the following.

The photon Green's function is defined by
\begin{align}
    A(t) &= -i \expct{T a(t) a^\dag(0)}{0} -i \expct{T a^\dag(t) a(0)}{0}.
\end{align}
Let us first assume that the cavity field is coupled to a thermal environment \cite{wallsmilburn}. The time-dependence of $a(t)$ is then governed by the Langevin equation, $\dot{a} = (-i \Omega - \kappa/2) a + \sqrt{\kappa} a_{\rm in}$, where $\kappa$ is the cavity decay rate and $a_{\rm in}(t)$ is the fluctuating thermal mode, satisfying $\expcts{a_{\rm in}^\dag(\omega) a_{\rm in}(\omega')}{} = 2 \pi n_{\rm th}(\omega) \delta(\omega - \omega')$, where $n_{\rm th}(\omega)$ is the Bose distribution. Approximating $n_{\rm th}(\omega) \approx n_{\rm th}(\Omega) =: n_\ph$, one finds for $n_\ph \gg 1$,
\begin{align}
    A(\omega) = - i \sum_{\eta = \pm} \frac{n_\ph \kappa}{(\kappa/2)^2 + (\omega - \eta \Omega)^2}.
\end{align}
The same result formally applies if we assume that the cavity is coherently driven at its resonance frequency $\Omega$, but in that case, $\kappa$ should be interpreted as the linewidth of the input field and $n_\ph$ is proportional to the number of photons in the cavity. For $\kappa = 0$, $A(\omega)$ consists of two delta peaks at $\omega = \pm \Omega$. A nonzero linewidth turns the latter into Lorentzians of width $\kappa$.

Finally, $G_{ij}(t)$ is the unperturbed Green's function of the CR. Defining $G_d(x,x',t) = -i \expcts{T d(x,t) d^\dag(x',0)}{0}$, one finds
\begin{align}\label{eq:G}
    G_{22}(t) &= G_{33}(t) = G_d(0,0,t) - G_d(0,0,-t), \\
    G_{23}(t) &= - G_{32}(-t) = -i G_d(0,L,t) - i G_d(L,0,-t). \notag
\end{align}
The Fourier transform of the Green's function $G_d(x,x',t)$ has the standard form, $G_d(x,x',\omega) = \frac{1}{2L} \sum_{k} \theta(k) e^{i k (x-x')}/[\omega - \epsilon(k) + i 0^+]$, where we took into account that $\sgn[\epsilon(k)] = 1$ due to the gapped spectrum.

Equations (\ref{eq:pert_D2})-(\ref{eq:G}) can be used to obtain the MBS Green's function,
\begin{align}\label{eq:Dgamma2}
    D_{22}(t)
&=
    \int \frac{d\omega}{2\pi} \frac{  e^{- i \omega t} \left[  \frac{\omega}{2} - \Sigma_{22}(\omega) \right] }{\left[\frac{\omega}{2} - \Sigma_{22}(\omega)\right]^2 + \Sigma_{23}(\omega)\Sigma_{23}(-\omega)}.
\end{align}
The residue theorem states that this Fourier transform at positive $t$ should be determined by the poles of the integrand in the complex lower half plane. For $g_c = 0$, there is a single pole at $\omega = 0$. For small $g_c$, the pole is split and shifted by an amount proportional to $g_c^2$. Since $\Sigma_{ij}(\omega)$ are regular functions near $\omega = 0$, we can replace $\Sigma_{ij}(\omega)$ by $\Sigma_{ij}(0)$. These functions are given by $\Sigma_{22}(0) = 0$ and
\begin{align}\label{eq:Sigma23}
    \Sigma_{23}(0) &=
    - i n_\ph g_c^2 m L \sum_{\eta=\pm}
    \frac{1}{\sqrt{\phi_\eta} \sin(\sqrt{\phi_\eta})},
\end{align}
where we defined $\phi_\pm = (\pm \Omega + \mu + i \kappa/2)/\epsilon_L$ and $\epsilon_L = (2 m L^2)^{-1}$ is proportional to the energy of the lowest excited state in the CR. Eventually, one finds,
\begin{align}\label{eq:Dgamma2_t}
    D_{22}(t>0) &= -i e^{-2 |\Gamma_R| t} e^{-2 i \sgn(\Gamma_R) \Omega_R t}, \notag \\
    D_{23}(t>0) &= \sgn(\Gamma_R) e^{-2 |\Gamma_R| t} e^{-2 i \sgn(\Gamma_R) \Omega_R t},
\end{align}
where $\Gamma_R = \Re[\Sigma_{23}(0)]$ and $\Omega_R = \Im[\Sigma_{23}(0)]$ denote the real and imaginary parts of the self-energy, respectively. The Green's functions thus display damped Rabi oscillations of the quantum state of the MBSs. The frequency $|\Omega_R|$ and damping rate $|\Gamma_R|$ are determined by the relation between the energy scales $\epsilon_L$, $\Omega$, $\kappa$, and $\mu$. For $\Gamma_R = 0$, the dynamics of the MBSs $\gamma_{2}$ and $\gamma_{3}$ according to Eq.~(\ref{eq:Dgamma2_t}),
\begin{align}\label{eq:Dgamma2_t_undamped}
    D_{22}(t>0,\Gamma_R = 0) &= -i \cos(2 \Omega_R t), \notag \\
    D_{23}(t>0,\Gamma_R = 0) &= -i \sin(2 \Omega_R t),
\end{align}
coincides with the prediction of the effective Hamiltonian (\ref{eq:V}) for $\varepsilon = 2 \Omega_R$. Therefore, Eq.~(\ref{eq:V}) can be regarded as the effective low-energy Hamiltonian which governs the time-evolution of the MBSs $\gamma_2$ and $\gamma_3$ if $|\Gamma_R| \ll |\Omega_R|$. For weak damping, the single-qubit rotations (\ref{eq:Uepsilon}) can thus be performed by driving the system for a finite time with a microwave frequency.

\begin{figure}[t]
    \includegraphics[width=\columnwidth]{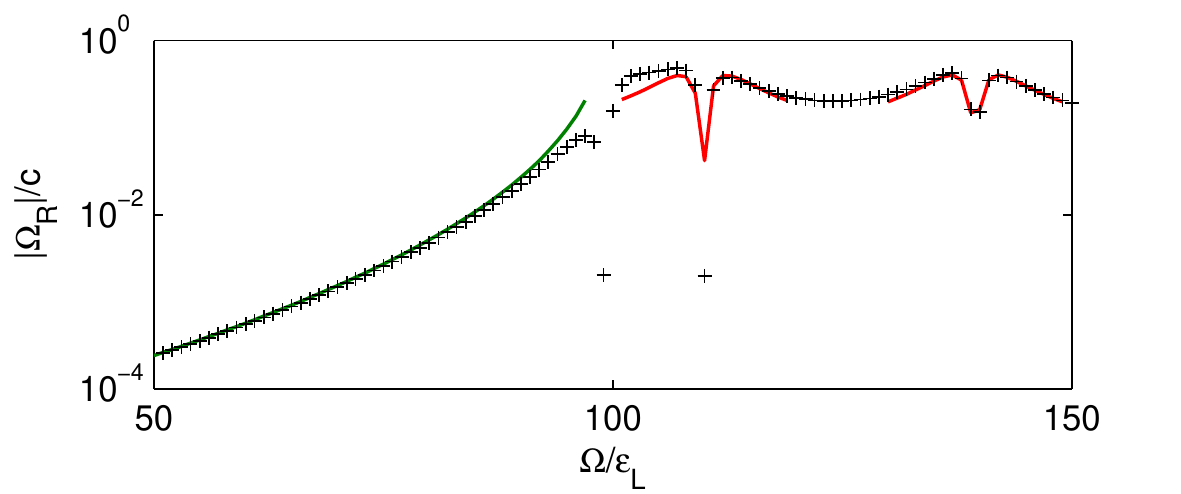}  \\
    \includegraphics[width=\columnwidth]{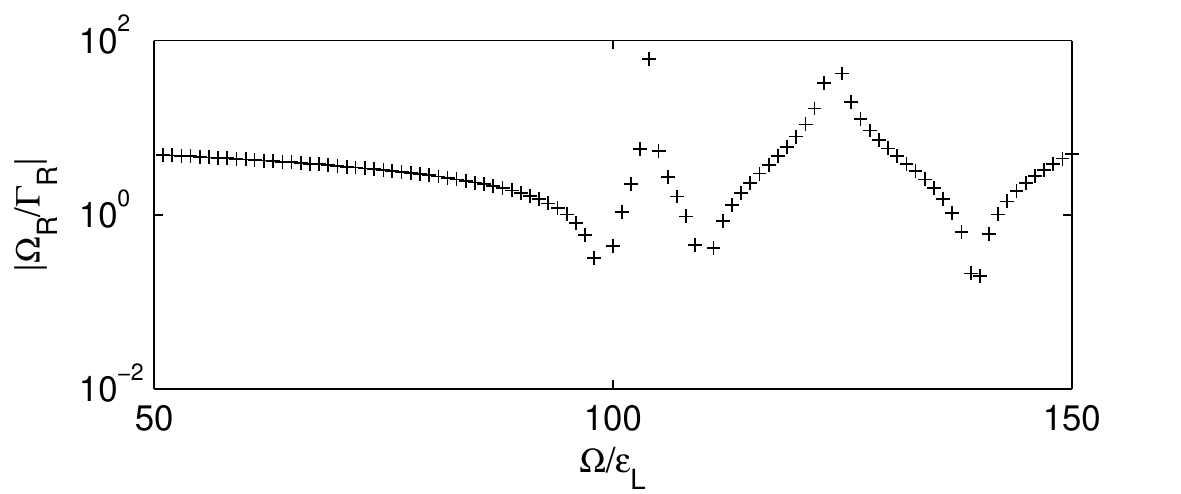}
    \centering
    \caption{(Color online) \emph{Upper panel:} Rabi frequency $|\Omega_R|$ in units of $c = n_\ph g_c^2/L$ for $\mu = -100 \epsilon_L$ and $\kappa = 5 \epsilon_L$, where $\epsilon_L = (2 m L^2)^{-1}$. A large photon linewidth $\kappa$ has been chosen to highlight the essential features. The crosses denote the Rabi frequency and damping determined numerically from Eq.~(\ref{eq:Dgamma2}). Solid green and red lines correspond to the solutions for the limits $\Omega < |\mu|$, see Eq.~(\ref{eq:GammaOmega_gapped}), and $\Omega > |\mu|$, see Eq.~(\ref{eq:GammaOmega_metallic}), respectively. \emph{Lower panel:} The ratio between Rabi frequency and damping, $\Omega_R/\Gamma_R$, determines the fidelity of qubit rotations.}
    \label{fig:plots}
\end{figure}

Since the topological protection of the MBSs relies on a large length of the CR, we shall assume $\epsilon_L \ll |\mu|, \Omega$. First, let us focus on the regime $\kappa \ll \Omega < |\mu|$, where the photon energy is insufficient to overcome the band gap, and only virtual tunneling processes are allowed. In this case,
\begin{align}\label{eq:GammaOmega_gapped}
|\Omega_R| &=
    \frac{n_\ph g_c^2 }{\sqrt{|\mu||\Omega + \mu|}\xi}\exp\left[- \sqrt{1 - \Omega/|\mu|}  \frac{L}{\xi} \right] \notag \\
\left|\frac{\Omega_R}{\Gamma_R}\right| &= \frac{4 \sqrt{|\Omega + \mu| \epsilon_L}}{\kappa}
\end{align}

These functions are plotted in Fig.~\ref{fig:plots}. The Rabi frequency is, as expected, exponentially suppressed in the length of the CR. However, as the photon frequency $\Omega$ approaches the critical value $|\mu|$, the prefactor $\sqrt{1 - \Omega/|\mu|} < 1$ leads to a significant increase of $\Omega_R$. The damping rate $\Gamma_R$ is determined by the photon linewidth $\kappa$. According to Eq.~(\ref{eq:Dgamma2_t_undamped}), a bit flip operation $\sigma_x$ can be achieved by rotating the qubit state for a time $t_* = \pi/(4 \Omega_R)$. In the presence of damping, the fidelity of such an operation can be estimated as
\begin{align}
    \mathcal{F} = e^{-2 |\Gamma_R| t_*} = \exp\left[-(\pi/2) |\Gamma_R/\Omega_R|\right].
\end{align}

Next, we discuss the case $\Omega > |\mu|$. In this regime, the photon field can excite real electrons from the MBSs into the CR. One still finds $\Sigma_{22}(0) = 0$, but the self-energy $\Sigma_{23}(0)$ now depends sensitively on the level structure of the CR. For $|\mu| \ll \kappa \ll \Omega$, the sine function in the denominator of Eq.~(\ref{eq:Sigma23}) causes Lorentzian resonances whenever the photon frequency matches an eigenenergy of the CR. For $\Omega \approx n^2 \pi^2 \epsilon_L - \mu$,
\begin{align}\label{eq:GammaOmega_metallic}
    |\Omega_R| &= \frac{n_\ph g_c^2}{L} \frac{|\Omega + \mu - n^2 \pi^2 \epsilon_L|}{(\Omega + \mu - n^2 \pi^2 \epsilon_L)^2 + (\kappa/2)^2}, \notag \\
    \left|\frac{\Omega_R}{\Gamma_R}\right| &= \frac{|\Omega + \mu - n^2 \pi^2 \epsilon_L|}{\kappa/2}
\end{align}
The resonances are shown in Fig.~\ref{fig:plots}. Finally, let us remark that similar Lorentzian resonances in $\Gamma_R$ and $\Omega_R$ as a function of $\Omega$ also arise in the limit of a very short CR, where one can replace the CR with a single fermionic level at energy $-\mu > 0$.

\emph{Conclusion.} We have shown that photon-assisted tunneling can have a strong impact on the coupling between two adjacent Majorana bound states (MBSs) which are separated by a gapped region. Absorption and emission of photons cause Rabi oscillations of the MBSs with a frequency that depends sensitively on the difference between the photon frequency and the gap width. Damping of the Rabi oscillations is caused by a nonzero photon linewidth. For subgap photon frequencies $\Omega<|\mu|$, the Rabi frequency is a monotonic function of $\Omega$ and increases exponentially for $\Omega \to |\mu|$. Such Rabi oscillations can be useful when applied to Majorana-based qubits because, complemented by braiding, they allow for the implementation of arbitrary single-qubit rotations.

\acknowledgments

This work was financially supported by the Swiss NSF
and the NCCR Quantum Science and Technology.

\bibliography{paper}

\end{document}